\documentclass[conference,twocolumn]{IEEEtran}
\ifCLASSINFOpdf
\else
\fi
\usepackage{amsfonts,amsmath,amssymb}
\usepackage{mathtools}
\usepackage{mdwmath}
\usepackage{cuted}
\usepackage{mdwtab}
\usepackage{amsthm}           % for designing theorems
\usepackage{bm}               % provides bold math \bm{} command (vectors)
\usepackage{units}            % for nicely typset units \unit[10]{kHz}
\usepackage{graphicx}
\usepackage{epstopdf}
\usepackage[linesnumbered,ruled,vlined]{algorithm2e}

\usepackage{url}
\usepackage{paralist}
\usepackage{color}
\usepackage{authblk}
\usepackage{cleveref}
\usepackage{float}
\usepackage[caption = false]{subfig}
\usepackage[nolist,nohyperlinks]{acronym} 
\newcommand{\Fig}[1]{Fig.~\ref{fig:#1}}


\begin{acronym}[NC-OFDM]
\acro{3gpp}[3GPP]{3\textsuperscript{rd} Generation Partnership Program}
\acro{5g}[5G]{Fifth Generation}
\acro{Adam}{Adaptive Moment Optimisation}
\acro{ap}[AP]{Access Point}
\acro{atc}[ATC]{Air Traffic Control}
\acro{bs}[BS]{Base Station}
\acro{bpp}[BPP]{Binomial Point Process}
\acro{cc}[C\&C]{Command \& Control}
\acro{cdf}[CDF]{cumulative distribution function}
\acro{cdma}[CDMA]{Code Division Multiple Access}
\acro{cdr}[CD\&R]{Conflict Detection \& Resolution}
\acro{cfo}[CFO]{Carrier Frequency Offset}
\acro{comreg}[ComReg]{Commission for Communications Regulation}
\acro{csi}[CSI]{Channel State Information}
\acro{easa}[EASA]{European Aviation Safety Agency}
\acro{faa}[FAA]{Federal Aviation Administration}
\acro{fso}[FSO]{Free-Space Optical}
\acro{geo}[GEO]{Geosynchronous Equatorial Orbit}
\acro{gps}[GPS]{Global Positioning System}
\acro{gs}[GS]{Ground Station}
\acro{hap}[HAP]{High Altitude Platform}
\acro{iot}[IoT]{Internet of Things}
\acro{kpi}[KPI]{Key Performance Indicator}
\acro{lap}[LAP]{Low Altitude Platform}
\acro{leo}[LEO]{Low Earth Orbit}
\acro{los}[LoS]{Line-of-Sight}
\acro{lte}[LTE]{Long Term Evolution}
\acro{mac}[MAC]{Media Access Control}
\acro{mcp}[MCP]{Matern Cluster Process}
\acro{mc}[MC]{Monte Carlo}
\acro{mimo}[MIMO]{Multiple Input Multiple Output}
\acro{mip}[MIP]{Mixed-Integer Programming}
\acro{mm}[MM]{Mapping Mechanism}
\acro{ml}[ML]{machine learning}
\acro{mno}[MNO]{Mobile Network Operator}
\acro{nlos}[NLoS]{non-Line-of-Sight}
\acro{nn}[NN]{Neural Network}
\acro{ofdma}[OFDMA]{Orthogonal Frequency Division Multiple Access}
\acro{oot}[OOT]{Out-of-Tree}
\acro{osi}[OSI]{Open Systems Interconnection}
\acro{otdoa}[OTDoA]{Observed Time Difference of Arrival}
\acro{ott}[OTT]{Over-The-Top}
\acro{pv}[PV]{photo-voltaic}
\acro{pdf}[pdf]{probability density function}
\acro{ppp}[PPP]{Poisson Point Process}
\acro{qos}[QoS]{Quality of Service}
\acro{rc}[RC]{Remote Control}
\acro{rl}[RL]{Reinforcement Learning}
\acro{rss}[RSS]{Received Signal Strength}
\acro{se}[SE]{Spectral Efficiency}
\acro{sir}[SIR]{Signal-to-Interference Ratio}
\acro{sinr}[SINR]{Signal-to-Interference-and-Noise Ratio}
\acro{snr}[SNR]{Signal-to-Noise Ratio}
\acro{uav}[UAV]{Unmanned Aerial Vehicle}
\acro{ue}[UE]{User Equipment}
\acro{ula}[ULA]{Uniform Linear Array}
\acro{urllc}[URLLC]{Ultra-Reliable Low Latency Communication}
\acro{wsn}[WSN]{Wireless Sensor Network}
\acro{wsn}[WSN]{Wireless Sensor Network}
\acro{rv}[RV]{random variable}
\acro{ppp}[PPP]{Poisson point process}
\acro{pgfl}[PGFL]{point generation functional}
\acro{pdf}[PDF]{probability density function}
\end{acronym}

\begin{document}
\title{Experimental Evaluation of a UAV User QoS from a Two-Tier 3.6GHz Spectrum Network}

\author{Boris Galkin$^1$,
        Erika Fonseca$^1$,
        Gavin Lee$^2$,
        Conor Duff$^2$,\\
        Marvin Kelly$^2$,
        Edward Emmanuel$^3$,
        and~Ivana Dusparic$^1$
%\thanks{This material is based upon works supported by the Science Foundation
%Ireland under Grants No. 10/IN.1/I3007 and 14/US/I3110. B. Galkin, J. Kibi\l{}da, and L. DaSilva are with %CONNECT, Trinity College Dublin, Ireland, email:  \{galkinb,kibildj,dasilval\}@tcd.ie.}% <-this % stops a space
}

\affil{$^1$ CONNECT- Trinity College Dublin, Ireland \\
        $^2$ Dense Air Ltd.\\
        $^3$ Smart Docklands, Dublin City Council\\
\textit{E-mail: \{galkinb,fonsecae,duspari\}@tcd.ie, \{glee,cduff,mkelly\}@denseair.net, edward@smartdocklands.ie}}

\maketitle

\begin{abstract}
%Providing omnipresent cellular connectivity to \acp{uav} is very challenging due to the dominance of \ac{los} interference and small gains from down-tilted \ac{bs} antennas.
Unmanned Aerial Vehicle (UAV) technology is becoming increasingly used in a variety of applications such as video surveillance and deliveries. To enable safe and efficient use of UAVs, the devices will need to be connected into cellular networks. Existing research on UAV cellular connectivity shows that UAVs encounter significant issues with existing networks, such as strong interference and antenna misalignment. In this work, we \textcolor{black}{perform a novel measurement campaign of the} performance of a UAV user when it connects to an experimental two-tier cellular network in two different areas of Dublin city's Smart Docklands, which includes massive MIMO macrocells and wirelessly-backhauled small cells. We measure Reference Signal Received Power (RSRP), Reference Signal Received Quality (RSRQ), Signal to Interference and Noise Ratio (SINR), the downlink throughput, \textcolor{black}{and the small cell handover rate}. Our results show that increasing the UAV height reduces the performance in both tiers, due to issues such as antenna misalignment. The small cell tier, however, can maintain relatively stable performance across the entire range of UAV heights, suggesting that UAV users can successfully connect to small cells during their flight. \textcolor{black}{Furthermore, we demonstrate that while the UAV handover rate significantly fluctuates at different heights, the overall observed handover rates are very low. Our results highlight the potential for small cells in urban areas to provide connectivity to UAVs.}
\end{abstract}


\section{Introduction}

Unmanned Aerial Vehicles (UAVs) are a technology that today is used widely for purposes such as photography, video recording and surveying \cite{8660516}. However, the scope of future applications is rapidly growing to include search and rescue, medical supply delivery \cite{transplant}, home delivery for retail, entertainment, agriculture as well as construction, increasing the critical need for reliable communications.
 
Today's applications require the UAV to have a reliable and fast wireless data connection to its pilot – most utilise the ISM band for a point-to-point link \textcolor{black}{via a remote-controller}. Current drone flight ranges are constrained by the distance of a pilot to a drone. This constraint is necessary to ensure visual control and to maintain the wireless connection for control; even with First Person View (FPV) flight, the range limitation is still enforced for public health and safety.
 
For future applications to become reality UAVs will have to utilise cellular networks via individual SIM cards. 5G connectivity is central to this as it will offer:
\begin{itemize}
\item Reduced latency providing sufficient communication and split-second control when flown out of sight (remote piloting) \cite{Ge_2019}.
\item Location information via GPS – tracking.
\item Cloud and/or edge computing to provide artificial intelligence for ‘swarm’ management.
\item Possibility to stream videos, for example, in city surveillance or big public events. 
\end{itemize}

For these reasons, cellular UAV connectivity has attracted significant research interest from the wireless community \cite{3GPP_2018,Mozaffari_2019,amer2019mobility,8756296,Galkin_2020}. However, existing research into LTE UAV connectivity has suggested that UAVs, due to their aerial nature, encounter significant issues when attempting to connect to the cellular network \cite{3GPP_2018,Mozaffari_2019}. UAVs have been shown to experience significant interference issues - when they fly above buildings they establish unobstructed Line-of-Sight (LOS) channels on distant interference sources \cite{qualcomm-sim}. UAVs have also been shown to suffer from Base Station (BS) antenna misalignment, as the downtilted BS antennas result in the main lobe of the antennas being directed to the ground, leaving the UAV to connect to the side-lobes that have weaker transmitted power \cite{Lin_2017}. Another experienced issue is the significantly increased number of handovers for the UAVs compared to the ground users. 

Some initial experimental trials of UAV-5G connectivity have been carried out by the wireless community. In \cite{Muzaffar_2020} the authors tested 4G and 5G connectivity in a rural environment and demonstrated that the UAV performed more handovers between 5G and 4G BSs as its height increased, due to it seeing distant 4G BSs at large heights. The authors of \cite{Seo_2020} tested a 5G connection in an urban area by moving a UAV directly up and down between heights of 0-100 meters above ground. Their results demonstrated that both downlink and uplink throughput decrease as the UAV height increases, with the downlink throughput dropping by more than 30\%. Note that all of the above experimental trials have considered only macrocell connectivity. \textcolor{black}{The wireless community has generally neglected the topic of UAV-small cell connectivity}; however, 5G networks will rely heavily on small cells for covering densely populated areas, which motivates this UAV experiment. %In this work, we carry out a set of experiments to measure the performance of a UAV user connecting into a two-tier network, containing both macrocell BSs as well as small cells. To our knowledge we are the first to carry out an experiment which explores how a UAV connects to and handovers between multiple small cells and a donor Macro using 3.6GHz spectrum.

In this research project we take measurements of a two-tier cellular network (macro and small cell) using 5G spectrum - 3.6GHz, in addition to implementing Massive MIMO technology on the macro layer. Whereas this spectrum range can cater for high speeds and greater user equipment (UE) connections, it has shorter range of radio frequency (RF) propagation. Therefore, small cells are deployed in the urban network to cover gaps in macrocell coverage. To our knowledge, we are the first to evaluate the performance of a UAV cellular UE when connecting to a two-tier network that includes small cells, on the 3.6GHz band. In the experiment, we evaluate the UAV connection in idle mode and dedicated mode as, depending on the UAV application, either mode can be used. A UE connected in idle mode does not transmit or receive any data in this state. It merely monitors the paging and broadcast channel so as to maintain connectivity. In the dedicated (also referred to as connected) mode, the UE is actively receiving or transmitting data. For example, a UAV which is carrying out a delivery mission and does not need a constant connection to its pilot would be in idle mode, whereas a UAV that is streaming video data would be in dedicated mode.

\section{Testbed Description}
\begin{figure}[t!]
\centering
	\includegraphics[width=.45\textwidth]{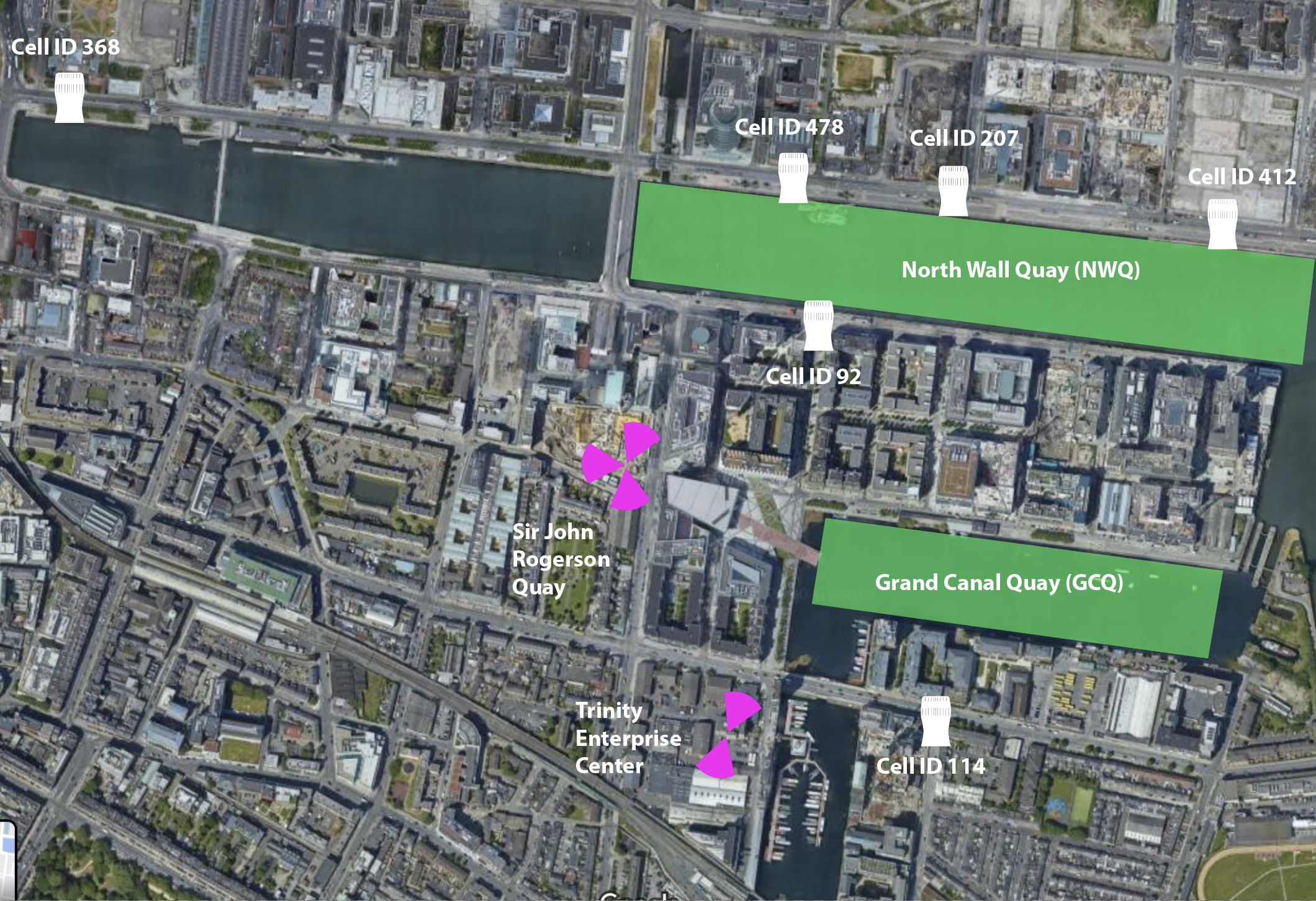}
%	\vspace{-30mm}
%	\vspace{15mm}
%	\vspace{-5mm}
	\caption{
 The Smart Docklands testbed. Macrocells are labelled in purple, small cells are denoted with white icons and measurement areas are denoted in green.
	}
	\label{fig:testbed}
		\vspace{-5mm}
\end{figure}

The experimental cellular network testbed is deployed in Dublin city’s designated Smart Docklands district in Ireland, shown in \Fig{testbed}. This district is situated around a river and a nearby waterway, with large open areas that are suitable for UAV deployment. The buildings in the area have heights between 20 and 80 meters. The testbed consists of three macrocells deployed on building rooftops, as well as a network of small cells deployed on lamp posts and traffic lights along the river. The macrocell labelled Trinity Enterprise Center in Fig. 1 will be the main macrocell that is tested in this experiment. This macrocell is a ZTE model ZXSDRB8300 with a 64 element antenna array Massive MIMO system. The lamppost small cells are AirSpeed model1250 with 4 antenna elements that apply 2x2 MIMO. Both the macrocell and the small cells operate on the B42 channel of the 3.6GHz 5G frequency band. The small cells operate on the frequency range 3410-3430MHz, while the macrocells operate at 3580-3600MHz. The macrocell antennas have a 15 degree Half-Power Beamwidth (HPBW), and the small cell antennas have 90 degree HPBW. The small cells wirelessly backhaul into the macrocell tier using the macrocell B42 channel, in effect acting as relays between the UE and the macrocell. The two-tier network has a hierarchical cell structure \cite{5GPPP_2018}, where the small cell tier have a higher connection priority than the macrocell tier. This means that a UE will prioritise connecting to the small cells even if it detects a stronger signal from a macrocell. Another aspect to the network on which we carry out this research is the Neutral Host nature of the network. Neutral Host is a communications model where each individual Radio Access Network (RAN) component can broadcast multiple Public Land Mobile Networks (PLMNs) – in essence a single small cell can communicate with UEs (SIM cards) belonging to multiple different network operators simultaneously. For future networks, and for future UAV applications, this will enable a single network infrastructure to provide a densified network for multiple UAV service providers. In this experiment, the handset which was connected to the UAV had a Dense Air SIM card.

\section{Experimental Procedure}
The experiment consisted of flying a handset which operated on Band 42 (3.6GHz) attached to a UAV through the urban area where the testbed is active at different heights above ground, while measuring the Key Performance Indicators (KPIs). For this experiment we used a DJI Matrice M300 UAV carrying a Google Pixel 3 handset. We designated two areas in the environment denoted as Grand Canal Quay (GCQ) and North Wall Quay (NWQ), where the water gives large open spaces suitable for flying but is flanked by buildings that create shading from the macrocells.
The UAV flew at a fixed height in a back-and-forth pattern at a speed of 8 m/s in the designated areas which were approximately 100 meters wide. This flight pattern was repeated at 10 meter increments for heights between 30 and 120 meters \textcolor{black}{(the legal flight ceiling)}. A single flight took between 12 and 15 minutes, and the UAV travelled approximately 2 kilometers. The handset operated in two modes: an idle mode where it recorded KPI related to channel quality such as Reference Signal Received Power (RSRP), Reference Signal Received Quality (RSRQ), and Signal to Interference and Noise Ratio (SINR), as well as in dedicated mode, where it was downloading data and was recording the downlink data throughput in addition to the KPIs. A sample set of raw data recorded during the experiment is available on our website \cite{SATORI_DATA}, and the full dataset can be provided upon request.

\begin{figure}[t!]
\centering
	\includegraphics[width=.45\textwidth]{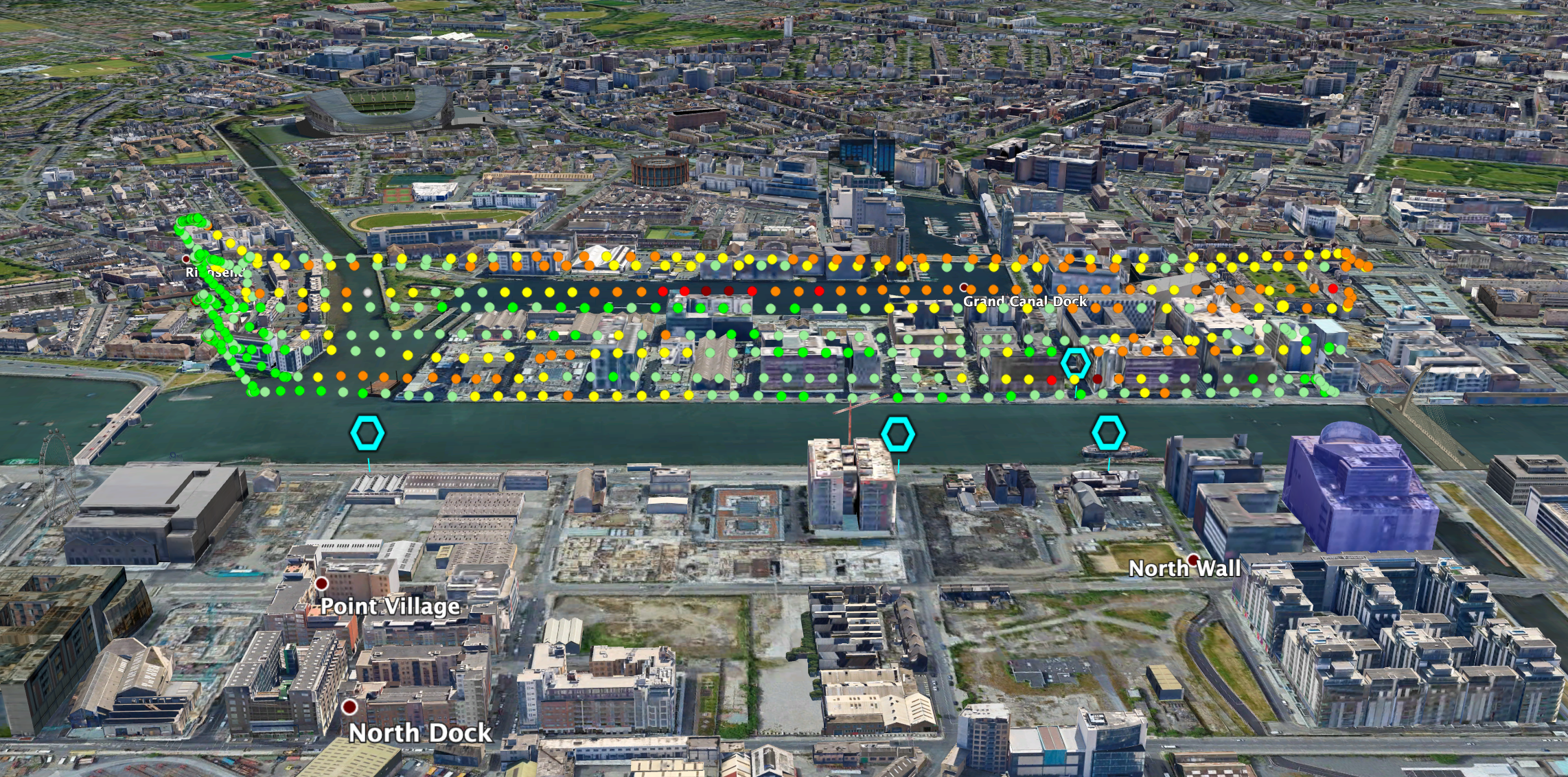}
%	\vspace{-30mm}
%	\vspace{15mm}
%	\vspace{-5mm}
	\caption{
 3D signal traces of the measured signal strength at different UAV heights along the NWQ flight area.
	}
	\label{fig:3Dtrace}
	%	\vspace{-5mm}
\end{figure}

\begin{figure}[t!]
\centering
	\includegraphics[width=.45\textwidth]{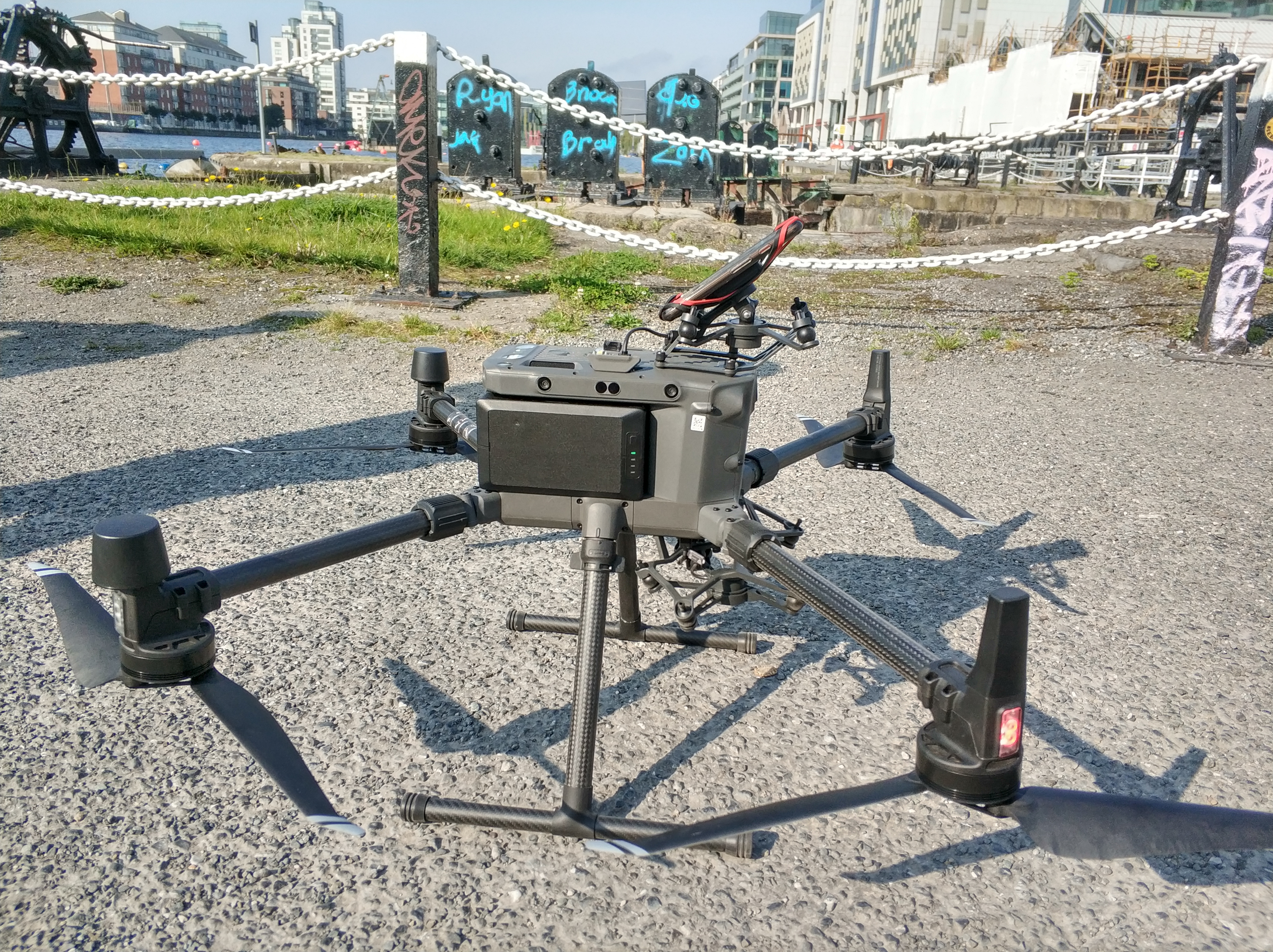}
%	\vspace{-30mm}
%	\vspace{15mm}
%	\vspace{-5mm}
	\caption{
 DJI Matrice drone with the top-mounted handset.
	}
	\label{fig:photo}
		\vspace{-5mm}
\end{figure}

\section{Experimental Results}
In this section we present the measurements obtained from our experiment. \textcolor{black}{For both locations we present the wireless channel KPIs in the form of the RSRP, RSRQ and SINR values, the downlink throughput in megabits per second (Mbps), as well as the handover rate and coverage percentage for the NWQ location}. We report that in the GCQ location the handset was always connected to the Trinity Enterprise Center macrocell \textcolor{black}{with no handovers occurring}, while at the NWQ location the handset was connected to the small cells, even though it could detect the macrocells at large heights. This is as designed, where the UE will consider the small cells as the highest priority layer due to the hierarchical cell structure. We first report and analyse the results in idle mode, for both test areas, followed by the results in dedicated mode. Following this, we report and discuss the handover behaviour in the NWQ location, as in this area the UAV experienced handovers between the multiple small cells.  

\subsection{Idle mode}
\begin{figure}[t!]
\centering
	\includegraphics[width=.45\textwidth]{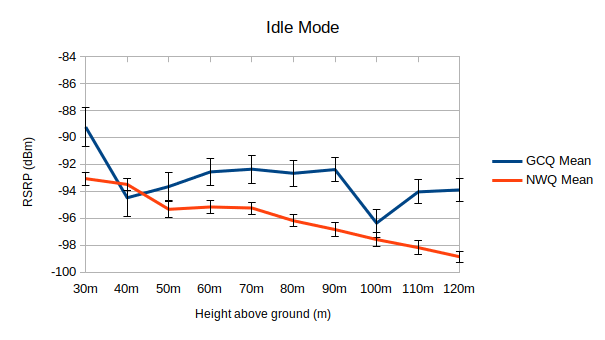}
%	\vspace{-30mm}
%	\vspace{15mm}
%	\vspace{-5mm}
	\caption{
 RSRP measurements at the GCQ and NWQ locations in idle mode, with 95\% confidence bounds.
	}
	\label{fig:RSRPidle}
	%	\vspace{-5mm}
\end{figure}

\begin{figure}[t!]
\centering
	\includegraphics[width=.45\textwidth]{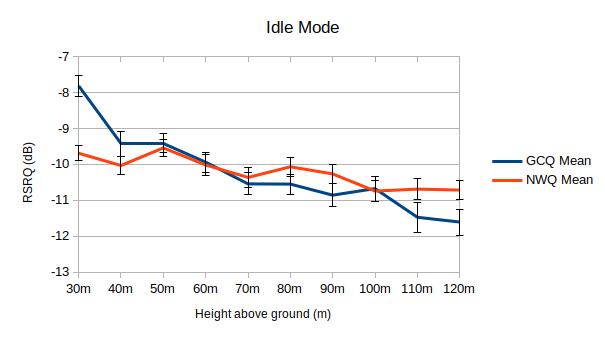}
%	\vspace{-30mm}
%	\vspace{15mm}
%	\vspace{-5mm}
	\caption{
  RSRQ measurements at the GCQ and NWQ locations in idle mode, with 95\% confidence bounds.
	}
	\label{fig:RSRQidle}
\end{figure}

\begin{figure}[t!]
\centering
	\includegraphics[width=.45\textwidth]{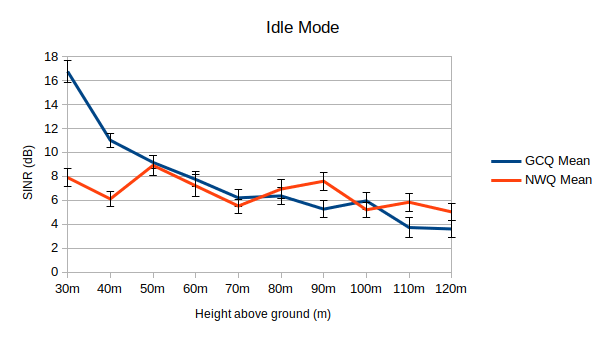}
%	\vspace{-30mm}
%	\vspace{15mm}
%	\vspace{-5mm}
	\caption{
 SINR measurements at the GCQ and NWQ locations in idle mode, with 95\% confidence bounds.
	}
	\label{fig:SINRidle}
%		\vspace{-5mm}
\end{figure}

\Cref{fig:RSRPidle,fig:RSRQidle,fig:SINRidle} show the measured RSRP, RSRQ and SINR values measured by the handset at different UAV heights in idle mode. We observe that increasing the UAV height has an overall negative impact on these KPIs, for both the macrocell and small cell connection. As the height of the UAV increases in the NWQ region it moves further away from the small cells, and so moves outside of their antenna radiation patterns. At the macrocell layer this behaviour is also predictable, as the antennas are tilted toward the ground area to serve ground UEs and are therefore not optimised for aerial connectivity. This prevents the UAV from receiving a strong signal when it is high above the ground. Furthermore, Massive MIMO makes use of strong multipath signals; when a UAV is high above ground it has one dominant LOS channel to the macrocell rather than several multipath channels, which reduces the benefits of Massive MIMO. 

The wireless community has established that when a UAV is operating at greater heights, the lack of building shadowing results in the UAV being able to receive strong interference signals from distant interference sources \cite{qualcomm-sim}. Our results confirm this effect. The average SINR measured at the GCQ drops by almost 15 dB as we increase the UAV height. It is interesting to note that in the NWQ location the UAV does not suffer as badly in the mean values. The average SINR fluctuates between 5 and 10dB across the entire range of heights, showing that the distance of the antenna influences the SINR considerably. A possible explanation is that the UAV is moving from side-lobe to side-lobe of the small cells, which provide a reliable connection to the UAV. Another possible explanation is that the small cells are deployed close to water, which reflects some of the signals towards the sky where the UAV operates.

	\vspace{5mm}

\vspace{-3mm}
\subsection{Dedicated mode}
\begin{figure}[t!]
\centering
	\includegraphics[width=.45\textwidth]{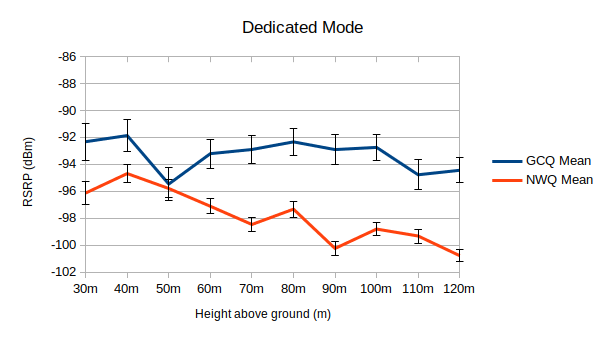}
%	\vspace{-30mm}
%	\vspace{15mm}
%	\vspace{-5mm}
	\caption{
 RSRP measurements at the GCQ and NWQ locations in dedicated mode, with 95\% confidence bounds.
	}
	\label{fig:RSRPdedicated}
%		\vspace{-5mm}
\end{figure}

\begin{figure}[t!]
\centering
	\includegraphics[width=.45\textwidth]{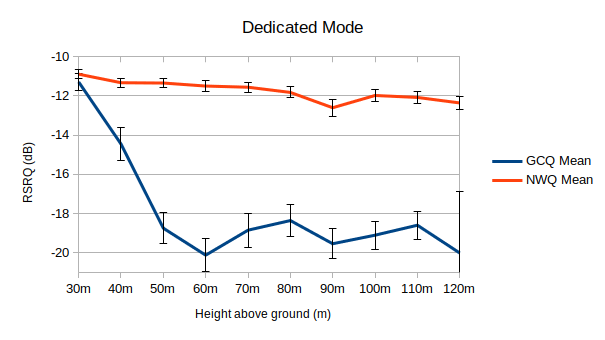}
%	\vspace{-30mm}
%	\vspace{15mm}
%	\vspace{-5mm}
	\caption{
 RSRQ measurements at the GCQ and NWQ locations in dedicated mode, with 95\% confidence bounds.
	}
	\label{fig:RSRQdedicated}
	%	\vspace{-5mm}
\end{figure}

\begin{figure}[t!]
\centering
	\includegraphics[width=.45\textwidth]{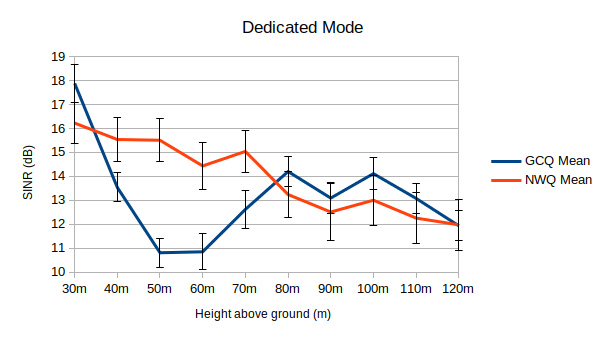}
%	\vspace{-30mm}
%	\vspace{15mm}
%	\vspace{-5mm}
	\caption{
SINR measurements at the GCQ and NWQ locations in dedicated mode, with 95\% confidence bounds.
	}
	\label{fig:SINRdedicated}
%	\vspace{-5mm}
\end{figure}

\begin{figure}[t!]
\centering
	\includegraphics[width=.45\textwidth]{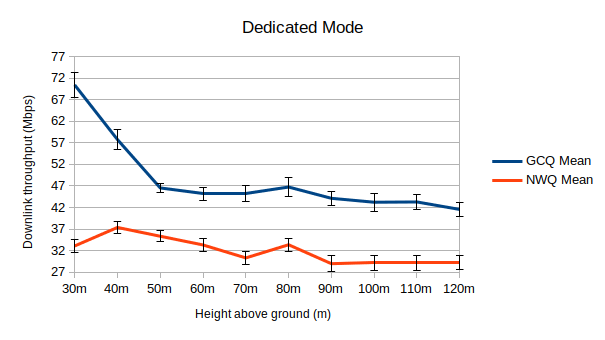}
%	\vspace{-30mm}
%	\vspace{15mm}
%	\vspace{-5mm}
	\caption{
 Downlink throughput measurements at the GCQ and NWQ locations in dedicated mode, with 95\% confidence bounds.
	}
	\label{fig:Throughput}
%		\vspace{-5mm}
\end{figure}

\Cref{fig:RSRPdedicated,fig:RSRQdedicated,fig:SINRdedicated,fig:Throughput} show the performance metrics measured in the dedicated mode when the handset is downloading data from the network. As in the idle mode, increasing UAV height appears to reduce the KPI values. We observe that the RSRP and RSRQ values for the NWQ location seem to be similar to those measured in idle mode, while for the GCQ location there appears to be a significant dip in the RSRQ performance in dedicated mode. It is interesting to note that both the GCQ and NWQ locations see a SINR increase between 5 and 10dB in the dedicated mode compared to the idle mode. This may be due to how the two tiers of cells apply MIMO in the dedicated mode.

Looking at the downlink throughput, we observe that the macrocell at the GCQ location provides a higher data rate than the small cells at the NWQ. Despite this, the GCQ data rate appears to suffer significantly at larger UAV heights, with the average data rate falling from 70Mbps at 30 meters to 40Mbps at 120 meters. Meanwhile, the small cells at NWQ appear to be able to provide a relatively consistent data rate of approximately 30-35Mbps on average, regardless of UAV height. The small cells show a more stable connection to the UAV UE because it does not fly as far away from the small cells as it does the macrocell, and the UAV always has a clear LOS to the small cells. The unstable connection directly to the macrocell is a result of the multipath environment with a UAV that is in motion. However, the macrocell on average provides a higher throughput than the small cells, due to Massive MIMO.

\subsection{\textcolor{black}{NWQ} handovers}
\begin{figure}[t!]
\centering
	\includegraphics[width=.45\textwidth]{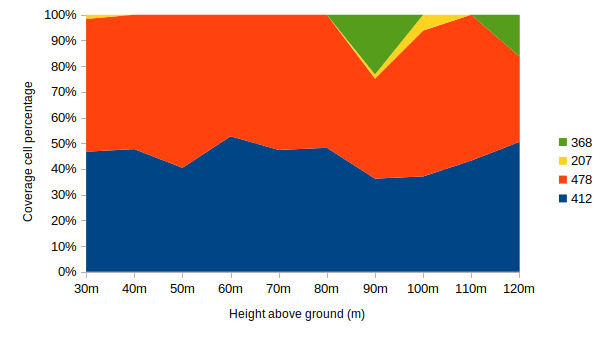}
%	\vspace{-30mm}
%	\vspace{15mm}
%	\vspace{-5mm}
	\caption{
Percentage of time the handset is connected to a given small cell.
	}
	\label{fig:coverage}
		\vspace{-5mm}
\end{figure}

\textcolor{black}{In the NWQ area the UAV handset connects to the 4 small cells along the river (cell IDs 368, 207, 478 and 412)}. The handset experiences handovers between the small cells, this handover behaviour changes as we increase the UAV height. \Fig{coverage} shows what percentage of the time the handset is connected to one of the \textcolor{black}{4} small cells for a given height. We see that at most heights the handset never connected to cell 368, the small cell approximately 1 kilometer west of the NWQ area; only above 80 meters does it begin to connect to it. This happens because at large heights the shape of the coverage areas created by the small cells change, as the UAV increases its distance to the terrestrial cells, while interacting with the antenna side-lobes. This result also shows that the distance to a cell is not as much a deciding factor for UAV UEs as it is for ground UEs; because of the antenna sidelobes a UAV may receive a stronger signal from a more distant cell than a closer one, and will connect accordingly. %We also note from our results that the UAV never connected to cell 92 on the south bank of the river. This may also be caused by antenna misalignment on cell 92, which results in the UAV receiving a stronger signal from the other cells.

\begin{figure}[t!]
\centering
	\includegraphics[width=.45\textwidth]{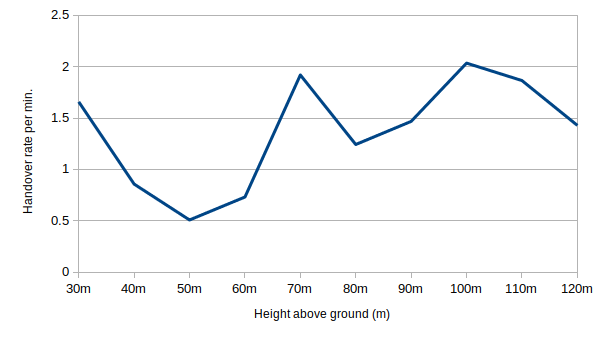}
%	\vspace{-30mm}
%	\vspace{15mm}
%	\vspace{-5mm}
	\caption{
 Handover rate at various UAV heights. 
	}
	
	\label{fig:handovers}
		\vspace{-5mm}
\end{figure}

\Fig{handovers} shows the average handover rate that occurs at each height over the NWQ. We observe that the handover rate fluctuates dramatically across different heights, \textcolor{black}{with an initial drop in the handover rate, followed by a rise, such that the handover rate at the lowest and highest tested heights appears to be almost the same}. We attribute this to a number of factors. First, we expect that at different heights the received powers from the small cell sidelobes change, which shapes the areas covered by those cells. Second, the water may reflect the signals from the small cells that are near the riverbank, which creates a rich multipath environment for the small cell signals. At large heights the handset is able to identify the distant macrocells, but always connects to the small cells in the area. What this means is that the rate of handovers fluctuates not because the handset begins connecting to new, distant macrocells, but because it \textcolor{black}{connects back and forth} between the same set of small cells. \textcolor{black}{It is interesting to note that although the handover rate quadruples as we move from 50 meters to 100 meters, the overall handover rates that occur are quite low. Despite the dense deployment of the small cells and the UAV moving at a speed comparable to a car, it has experienced at most 1 handover every 30 seconds. We attribute this to the UAV receiving strong, unobstructed signals from all of the nearby small cells, which reduces the probability of it detecting a sufficiently large signal difference between its connected cell and a neighbouring cell and prevents it from triggering a handover often \cite{ericssonttt}.}%It is interesting to note that this dramatic variation in handovers does not appear to negatively affect the throughput observed by the handset at the corresponding heights. 

%\vspace{-3mm}

\subsection{\textcolor{black}{Results Discussion}}
\textcolor{black}{The results of our experiments provide several interesting observations regarding both macrocell as well as small cell cellular service for UAV UEs. For the macrocell tier we demonstrated that increasing the UAV height above ground significantly deteriorates the channel quality and the resulting downlink throughput. This observation is in line with existing research on UAV connectivity. It is significant to note, however, that our observations were made for a macrocell using massive MIMO antenna technology, unlike many prior works which have instead considered conventional LTE macrocells with fixed antenna radiation patterns \cite{3GPP_2018,Mozaffari_2019}. Prior works such as \cite{Geraci_2018} have suggested that massive MIMO will be a key enabler for cellular UAV connectivity, due to benefits such as interference mitigation and beamforming. Our results, however, suggest that even when equipped with massive MIMO the macrocell tier experiences significant performance degradation, and that additional network planning on the part of the network operator would be required to address this performance loss.}

\textcolor{black}{
Our results have also shown the capabilities of the small cell tier for UAV connectivity. Due to their aerial nature and ability to connect to a number of distant macrocells with unobstructed channels, the wireless community tends to assume that UAVs will be served by macrocells, even in environments where small cells may be available \cite{Galkin_2020}. Our experimental trials have demonstrated that not only are UAV UEs able to successfully connect to small cells even at large heights (up to the 120 meter flight ceiling), but that the downlink data rate delivered by the small cells remains quite consistent across the range of UAV heights. Our experiments have also highlighted that UAVs do not encounter significant handover rates when connected to the small cells, even at large heights. Due to the strong received signals from multiple small cells at the UAV it does not trigger handover events very often. This suggests that even rapidly-moving UAVs may be able to benefit from small cell connectivity at large heights above ground.
}

\section{Conclusion}

In this paper we have performed real-world measurements of the performance of an experimental two-tier cellular network testbed, using a handset mounted on a UAV. These measurements were taken to investigate how the UAV UE interacts with the cellular network, and whether the network is capable of delivering service to the UAV UE.
Our results have shown that the network performance deteriorates as the UAV increases its height, due to the effects of antenna radiation pattern misalignment, increased distance to the cells, as well as increased interference. This resulted in the macrocell throughput dropping by more than 30\% as the UAV moved from 30 meters above ground to 120 meters. We observed that the macrocell suffered more from this performance degradation than the small cells, which were able to provide a relatively consistent data throughput across all heights. \textcolor{black}{We also observed that the handset experienced a very small number of handovers from the small cells. Although the handover rate fluctuated dramatically for the different UAV heights, the range of values were still very low.}

\textcolor{black}{In our future work we intend to build on the observations made from this experimental data, to explore the ability of small cells to provide connectivity to UAV UEs in 5G and beyond networks.}
%It has been shown that the relay small cell network can densify the macro network. A massive MIMO system (such as in 5G) can provide an excellent backhaul connection for small cells and hence the small cells can provide consistent data throughput in poor macro connection areas, even for UAV users operating at large heights above ground.

%As already mentioned, the performance results at large heights may be influenced by the water reflections in the quay. Further experimentation is needed to examine this effect and its impact on UAV-cellular connectivity. 

\section{Acknowledgement}
%This material is based upon works supported by the Science Foundation Ireland under Grants No. 17/NSFC/5224, 16/SP/3804 and
%13/RC/2077. The authors would also like to thank Mr. Fergal McCarthy of Drone Services Ireland for his assistance.

This work was supported by a research grant from Science Foundation Ireland (SFI) and the National Natural Science Foundation Of China (NSFC) under the SFI-NSFC Partnership Programme Grant Number 17/NSFC/5224, as well as SFI Grants No. 16/SP/3804 and
13/RC/2077. The authors would also like to thank Mr. Fergal McCarthy of Drone Services Ireland for his assistance.

\ifCLASSOPTIONcaptionsoff
  \newpage
\fi



\bibliographystyle{./IEEEtran}
\bibliography{./IEEEabrv,./IEEEfull}
\end{document}